\documentclass[11pt]{scrartcl}
\usepackage[margin=2.5cm,a4paper,twoside]{geometry}

\usepackage{amssymb,amsmath,graphicx}
\usepackage{pgfplots,pgfplotstable}
\usepackage[ascii]{inputenc}
\pgfplotsset{compat=newest}
\pgfplotsset{
  log x ticks with fixed point/.style={
      xticklabel={
        \pgfkeys{/pgf/fpu=true}
        \pgfmathparse{exp(\tick)}%
        \pgfmathprintnumber[fixed relative, precision=3]{\pgfmathresult}
        \pgfkeys{/pgf/fpu=false}
      }
  }
}

\usepackage[english,french]{babel}
\usepackage{stmaryrd}

\usepackage{color}

\newcommand{\refdom}{\hat\Omega}
\newcommand{\trfdom}{\Omega}

\newcommand{\param}{\alpha}
\newcommand{\transform}{\psi}
\newcommand{\deform}{\phi}
\newcommand{\refdeform}{\hat{\phi}}
\newcommand{\eigenstrain}{U}
\newcommand{\energy}{{\mathcal E}}
\newcommand{\refenergy}{{\hat{\mathcal{E}}}}
\newcommand{\cost}{{\mathcal J}}

\newcommand{\edensity}{W}
\newcommand{\vara}{1}
\newcommand{\varb}{2}
\newcommand{\varboth}{{1,2}}
\renewcommand{\d}{\,\mathrm{d}}
\newcommand{\xref}{\hat{x}}
\newcommand{\D}{{D}}
\newcommand{\Dref}{\hat{D}}
\newcommand{\R}{\mathbb{R}}
\newcommand{\tr}{\mathrm{tr}}
\newcommand{\Id}{\mathrm{Id}}
\newcommand{\SO}{\mathrm{SO}}
\newcommand{\GPa}{\mathrm{\,GPa}}
\newcommand{\NiAl}{Ni$_{65}$Al$_{35}$}
\definecolor{vara}{rgb}{0.8,1.0,0.8}
\definecolor{varafull}{rgb}{0.1,0.6,0.1}
\definecolor{varb}{rgb}{0.8,0.8,1.0}
\definecolor{varbfull}{rgb}{0.1,0.1,0.6}
\definecolor{coord}{rgb}{0.7,0.0,0.0}
\definecolor{meas}{rgb}{0.5,0.5,0.5}

\newcommand{\cross}{$\boxslash$}
\newcommand{\criss}{$\boxbslash$}
\begin{document}

\selectlanguage{english}

\title{Geometry of martensite needles in shape memory alloys}
\author{Sergio Conti, Martin Lenz, Nora L\"uthen, \\ Martin Rumpf, Barbara Zwicknagl}

\maketitle

\begin{abstract}
We study the geometry of needle-shaped domains in shape-memory alloys.
Needle-shaped domains are  ubiquitously found in martensites around macroscopic interfaces between regions which are laminated in different directions, or close to macroscopic austenite/twinned-martensite interfaces. Their geometry results from the interplay of the local nonconvexity of the effective energy density with long-range (linear) interactions mediated  by  the elastic strain field, and is up to now poorly understood. 
We present a two-dimensional shape optimization model based on finite elasticity and discuss its numerical solution.
Our results indicate that the tapering profile of the needles can be understood within finite elasticity, but not with linearized elasticity. The resulting tapering and bending reproduce the main features of experimental observations on \NiAl.
\end{abstract}

\section{Introduction}
\label{sec:intro}

Shape-memory alloys couple complex macroscopic material behavior with specific microstructures \cite{salje1990phase,Bhatta}. 
Low-energy interfaces between different variants of martensite, or between martensite and austenite, are possible only under specific conditions, and only with a few  special orientations. Indeed, 
in the martensitic phase one often observes finely twinned laminates with a crystallographically-determined orientation.
Most experimental situations require deviations from those orientations and lead to complex structures, which involve curved or complex interfaces.

\begin{figure}[t]
\begin{center}
\includegraphics[width=\textwidth]{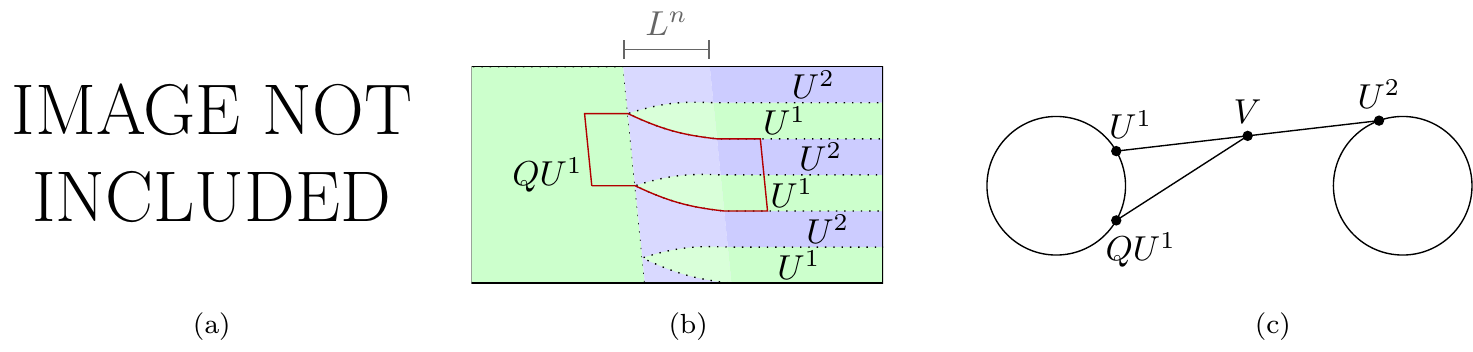}
\end{center}
\caption{(a) Experimental observation of needle-shaped domains via transmission electron microscopy in a thin sample of \NiAl.
Available under 
https://doi.org/10.1103/PhysRevB.64.144105. (b) Sketch of the  geometry in the reference configuration. The red line outlines the fundamental cell, which is repeated periodically in the vertical direction. (c) Sketch of the geometry in matrix space. The macroscopic deformation gradient $V$ is a weighted average of $\eigenstrain^\vara$ and $\eigenstrain^\varb$, and is itself compatible -- along a different direction -- with $Q\eigenstrain^\vara$ for some $Q\in\SO(2)$.}
\label{figureneedle1}
\end{figure}

One typical situation is the appearance of so-called ``needle-like'' domains \cite{salje1990phase,SaljeIshibashi1996,BoScKo01}, see Figure \ref{figureneedle1} for an illustration. 
They often appear close to macrointerfaces between austenite and finely twinned martensite or between regions where the martensite is twinned in different directions. Needle-like domains
are believed to be crucial for the macroscopic energetics and for the hysteresis of the phase transition, as they determine the cost of the transition state \cite{ZhangJames-2005,ZhangJamesMueller-2009,Zwicknagl2014}.
They are thin domains of one martensite variant, which taper approaching a tip.
The bending of the needle was related 
to its tapering \cite{BoScKo01}, but the tapering mechanism is up to now not understood. In particular it is not clear what determines the length scale of the needles. 

A related (and competing) microstructure is the so-called branching pattern, first studied in this context by Kohn and M\"uller \cite{KohnMueller-1994}, which can be seen as a laminate which refines close to the macrointerface. In some models branching extends down to scale zero, leading to asymptotically self-similar deformation patterns \cite{Conti00}. If interfacial energy penalizes interfaces with all orientations, however, branching is expected to stop at some point, and, close to the interface, to be replaced by a single-scale interpolation. The latter might correspond to needles, see for example \cite{ContiZwicknagl2016}.
The detailed geometry of branching patterns was studied in \cite{DondlHeerenRumpf2016}.
The role of needles can be further strengthened by evolutionary aspects, as the gain in energy made possible by branching patterns is not necessarily dynamically accessible to the material during the phase transformation. We do not investigate branching patterns further here.

The geometry of needles was studied analytically in a geometrically linear setting in
\cite{SaljeIshibashi1996,BoScKo01,BoullaySchryversBall2003}. 
This permitted in particular to find a simple relation between the bending of the thin needles and their tapering, but the tapering profile itself could not be predicted.
Numerical studies with geometrically linear models also did not lead to a stable prediction of the tapering profile \cite{muite-salman}.
There has been a considerable effort to reproduce the formation of needles numerically, with models based on nonlinear elasticity and many different discretizations schemes. 
Simulations based on a phase field method \cite{finel-et-al:10} have shown 
that a geometrically nonlinear model is needed for the modelling of polytwinned microstructures.
Needle formation, bending, tapering and branching have been observed in studies based on 
finite elements with a nonconvex energy density \cite{LiLuskin1999}. 
Needles close to a free surface have been studied with 
an atomistic discretization \cite{NovakBismayerSalje2002},
showing that they have a strong tendency to either retract or to transform into a laminate. 
 An analysis based on three-dimensional FEM with parametrized boundaries was presented, for the case of Cu-Al-Ni, in \cite{SeinerGlatzLanda2011}. This paper focused on energy concentration around the needle tip. The resulting needle geometry was qualitatively similar to the experimental one, but there remained significant differences \cite[Fig.~8 and 9]{SeinerGlatzLanda2011}.  
 A phase-field approach, also with a geometrically nonlinear model, has been presented in \cite{LevitasRoyPreston2013} and also shows bending and tapering of needle-like structures (see in particular Figure 3 there).  
 In \cite{BrFaSh19} the stored elastic energy was numerically minimized with respect to variation in the period of the twinned microstructure. 
 The formation of needles was also observed using a Fourier-space discretization of a viscoelastic model \cite{SalmanMuiteFinel2019}.
Whereas all these studies agree in showing that models based on nonlinear elasticity can predict the formation of needles which qualitatively resemble the observed ones, the detailed shape, and its dependence on the parameters of the problem, was not explored. 

In this work,  we  use a two-dimensional nonlinear elasticity model to show numerically that the tapering profile can be understood as a consequence of geometric nonlinearity. 
Our results indicate that the effective length of the needle is proportional to $1/\delta$, where $\delta$ is the order parameter in the bulk.  Correspondingly, a geometrically linear version of the model predicts an infinite length of the needles.  
For definiteness, we focus on needles completely in the martensitic phase of {\NiAl} and on the TEM observations reported in  \cite{BoScKo01}, but our model can be easily generalized to other materials.

\section{Model}
\label{sec:mod}
 We work with nonlinear elasticity, following \cite{BallJames87}, in two  spatial dimensions. 
We assume that there are two martensitic variants, with eigenstrains
\begin{equation}\label{eqdefeigenstrains}
\eigenstrain^\vara:=\begin{pmatrix}1 &\delta\\ 0&1\end{pmatrix} \text{ and }
\eigenstrain^\varb:=\begin{pmatrix}1 &-\delta\\ 0&1\end{pmatrix} 
\end{equation}
for some $\delta>0$; in the simulations we use $\delta=0.1$ which is a typical value for {\NiAl} (see \cite[Sect. IV/A]{BoScKo01}). We use, as customary, the austenite as reference configuration, scaled to have the same density as the martensite (so that $\eigenstrain^\vara$ and $\eigenstrain^\varb$ have unit determinant). The minimizers of the free-energy density are then the deformation gradients of the form $R\eigenstrain^\vara$ or $R\eigenstrain^\varb$, for any $R\in\SO(2)$. If there is a continuous interface between a region with deformation gradient $R\eigenstrain^\vara$ and one with deformation gradient $R'\eigenstrain^\varb$, then necessarily $R\eigenstrain^\vara\tau=R'\eigenstrain^\varb\tau$ for any tangent vector $\tau$. This equation has two  pairs of solutions for $\tau\in S^1$, the first one being  $\tau=\pm e_1$ (with $R'=R$); and the second one being $\tau=\pm e_2$ (with $R$, $R'$ related by $R(\delta,1)=R'(-\delta,1)$). 

We assume that the left part of the sample is in variant 
$\eigenstrain^\vara$, and the right part is in a mixture of variants $\eigenstrain^\vara$ and $\eigenstrain^\varb$
 (see Figure \ref{figureneedle1}(b)).
We denote by $\Omega^\vara$ (or $\Omega^\varb$) the part of the domain in which variant $\eigenstrain^\vara$ 
(or $\eigenstrain^\varb$) is used, and minimize the energy
\begin{equation}\label{eqdefetot}
 E[\deform,\Omega^\vara, \Omega^\varb] := \int_{\Omega^\vara} W^\vara(D\deform) dx + \int_{\Omega^\varb} W^\varb(D\deform ) dx ,
\end{equation}
where $\deform:\Omega^\vara\cup\Omega^\varb\to\R^2$ is the elastic deformation, and $W^i$ is the free-energy density of variant $i$.
We do not include a surface-energy term, proportional to the length of the interface between $\Omega^\vara$ and $\Omega^\varb$. This contribution is typically much smaller than the elastic energy. The surface energy is crucial in determining the length-scale of the $\eigenstrain^\vara$/$\eigenstrain^\varb$ mixture on the right of the interface, which determines the number of interfaces and hence their total length, an effect we do not study here. The bending and tapering of the needle instead have only a minor effect on the total length of the interface, whereas they have a significant effect on the long-range elastic incompatibility. Therefore surface energy is not crucial in this situation, and for simplicity we do not include it in the model.

We impose boundary conditions on the two domains $\Omega^\vara$, $\Omega^\varb$ corresponding to the geometry illustrated in Figure \ref{figureneedle1}(b), and determine both the elastic deformation $\phi$ and the two sets $\Omega^\vara$, $\Omega^\varb$ by minimizing the energy $E$, resulting in a shape-optimization problem. As we do not optimize over the topology, we opt for a reparametrization scheme (discussed below) for the optimization of the two shapes.
The domains $\Omega^\vara$, $\Omega^\varb$ are parametrized via low-order polynomials.

We assume that on the far right a periodic mixture of the $\eigenstrain^\vara$ and $\eigenstrain^\varb$ variants is present, with periodicity $H$, and volume fractions $\theta$ and $1-\theta$, $\theta\in (0,1)$. This is geometrically possible, since 
$\eigenstrain^\vara e_1=\eigenstrain^\varb e_1$. The average deformation gradient in this region is then $
V:=\theta \eigenstrain^\vara + (1-\theta) \eigenstrain^\varb = \Id + (2\theta-1)\delta e_1\otimes e_2$ and is
compatible to a different rotation of the $\eigenstrain^\vara$ variant,
$Q\eigenstrain^\vara e_2^{\delta \theta}=Ve_2^{\delta \theta}$, where
 $e_2^{\delta \theta}:= \frac{(-\delta \theta,1)}{\sqrt{1+\delta^2 \theta^2}}$.
 The rotation $Q$ can be easily computed in terms of $\delta$ and $\theta$.
We are interested in resolving the structure around this macrointerface, 
which is oriented along  $e_2^{\delta \theta}$,
assuming the periodicity is not changed. 

We use a generic polyconvex nonlinear elastic energy density with the cubic symmetry of the austenite \cite{Ka07},
\begin{equation}\label{eqdefendensity}
\edensity(F) := a_1 \, \tr(F^T F)^2 + a_2 \, \det(F^T F) - a_3 \, \log(\det(F^T F)) + a_4 \, \big((F^TF)_{11}^2+(F^TF)_{22}^2\big)\,,
\end{equation}
 and extend it to the two martensite variants by 
setting
$\edensity^\vara(F) := \edensity(F(\eigenstrain^\vara)^{-1})$, 
$\edensity^\varb(F) := \edensity(F(\eigenstrain^\varb)^{-1})$.
If $a_3=4a_1+a_2+2a_4$ and suitable inequalities hold, then $W$ is minimized on $\SO(2)$. The remaining three coefficients correspond to the three elasticity constants of a cubic material. We choose $c_{11}= 200 \GPa$, $c_{12} = 130\GPa$, $c_{44}=110 \GPa$, which are appropriate 
for the B2 austenite phase of {\NiAl},  see for example  \cite[Table I]{HuangNaumovRabe2004} and set $a_1=11.56\GPa$, $a_2=-17.44\GPa$, $a_3=10.04\GPa$, $a_4=-9.38\GPa$.

\section{Results}
\label{sec:results}
We have numerically optimized the total elastic energy $E[\deform,\Omega^\vara, \Omega^\varb]$ 
both in the deformation $\deform$ and in the shape of the domains $\Omega^\vara$ and $\Omega^\varb$. We use a finite-element discretization 
for the deformation and minimize it out for any fixed domain geometry. The resulting effective functional, which depends only on the domains, is then minimized in the shape of the needles, which is parametrized in terms of 
the total needle length $L^n$ and 
two quadratic curves $\gamma^t$ and $\gamma^b$ describing the top and bottom needle profiles in the reference configuration, respectively. Since we assume periodicity along the interface, only one needle needs to be resolved numerically.
For a detailed explanation of the concrete parametrization of the needle geometry and the shape optimization method we refer to Section \ref{sec:opt}.

\begin{figure}[ht]
\begin{center}
\includegraphics[width=0.98\textwidth]{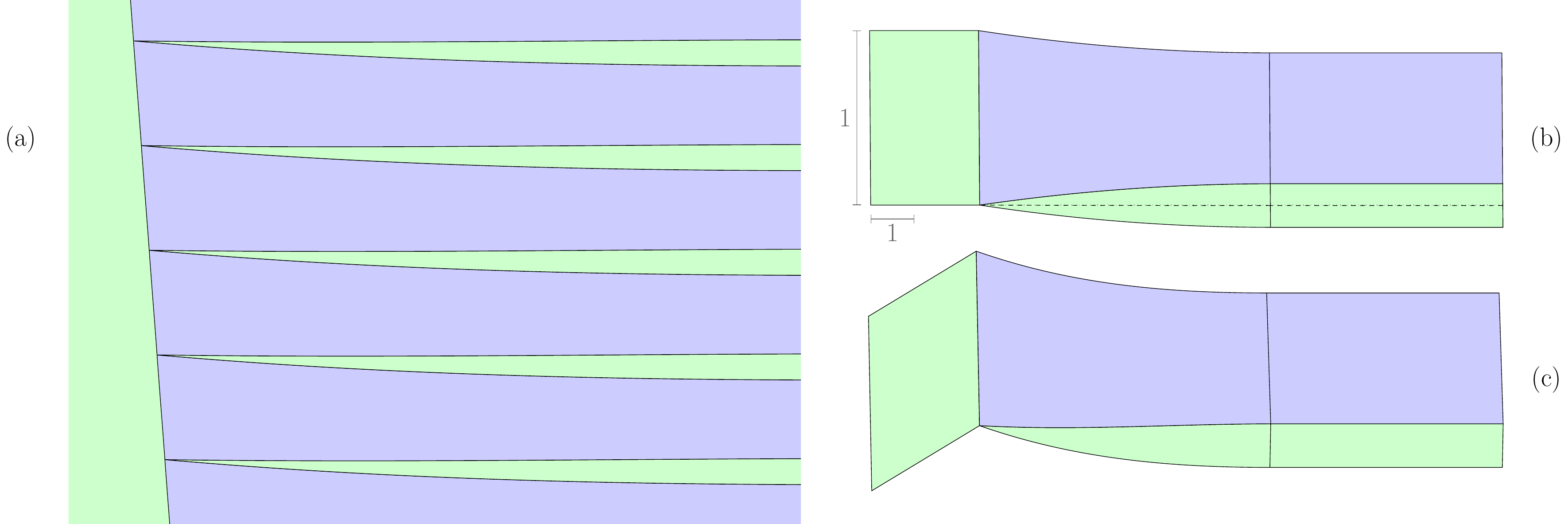}
\end{center}
\caption{
(a) Periodic needle pattern (in the deformed configuration) as obtained by numerical minimization of the functional (\ref{eqdefetot}) with $\delta = 0.1$ and $\theta = 0.25$, to be compared with the experimental observation in Figure~\ref{figureneedle1}(a).
We also show the needle shape in the reference (b)  and  deformed (c) configuration,
 scaled by a factor of 4 vertically in order to better illustrate the shape of the needle. The shape in the reference configuration is almost symmetric, but in the deformed configuration a substantial bending appears.}
\label{fig:result:stdpars}
\end{figure}

The resulting shape of the needles is illustrated in Figure~\ref{fig:result:stdpars}(a). The needle tapering naturally occurs on a length scale which is of the order of 5 to 10 needle spacings, in agreement with experiment (see Figure~\ref{figureneedle1}(a)). This tapering effect was not resolved in the geometrically linear model in \cite{BoScKo01} and is discussed in more detail below. The tapering in turn generates a bending of the needle, as apparent both in the numerical results of Figure~\ref{fig:result:stdpars}(a) and in the experimental image in Figure~\ref{figureneedle1}(a). In the reference configuration the bending of the needle is almost absent, and only a small asymmetry between the two boundaries is apparent  (see Figure~\ref{figureneedle1}(b) and discussion below). This confirms that the experimentally observed bending results from the fact that compatibility across a tapering needle requires a nontrivial rotation, as predicted by the geometrically linear analysis \cite{BoScKo01}. In the following we discuss the tapering length and the asymmetry in more detail.

\paragraph*{Needle length.}
Figure~\ref{fig:result:stdpars} shows the optimal needle shape obtained for $\delta=0.1$ and volume fraction $\theta=0.25$.
The right panel shows an enlargement of the shape of a single needle, which is almost, but not exactly, symmetric in the reference configuration (see below).

Figure~\ref{fig:nonlineardelta} shows that the general structure with an almost symmetric tapering in the reference configuration and a substantial bending in the deformed configuration is not specific to $\delta=0.1$. At the same time the effective tapering length $L^n$ (defined precisely via the formulas for $\gamma^t$ and $\gamma^b$ given in Section~\ref{sec:opt} below) depends strongly on $\delta$ (defined in (\ref{eqdefeigenstrains})). By scaling one easily sees that necessarily $L^n=Hf(\delta)$ for the present scale-invariant model. The right panel shows that the dependence is very well approximated by $L^n=c H/\delta$, predicting an infinite tapering length in the limit $\delta\to0$. This is precisely the limit in which linearized elasticity applies.

\begin{figure}[ht]
\begin{center}
\begin{tikzpicture}
\begin{scope}[scale=0.9]
\begin{loglogaxis}[height=6cm,width=6cm,log ticks with fixed point,xtick={0.05,0.1,0.2},ytick={4,6,8,10,12},axis x line*=bottom,axis y line*=left,axis line style={-latex},xlabel={$\quad\delta$},ylabel={$L^n$},xlabel style={at={(rel axis cs:1,0)}},ylabel style={at={(rel axis cs:0,1)},rotate=-90}]
\addplot[domain=0.04:0.25]{1.30086933e+01/(x/0.05)};
\addplot[mark=o, color=varafull, only marks] table {
0.05  1.30086933e+01
0.06  1.09477475e+01
0.07  9.43285913e+00
0.08  8.27479226e+00
0.09  7.36211212e+00
0.10  6.62474989e+00
0.11  6.01708848e+00
0.12  5.50779903e+00
0.13  5.07499862e+00
0.14  4.70268969e+00
0.15  4.37909826e+00
0.17  3.84436437e+00
0.20  3.24056723e+00
};
\addplot[color=varafull,only marks] table {
0.05  1.30086933e+01
0.10  6.62474989e+00
0.20  3.24056723e+00
};
\end{loglogaxis}
\end{scope}
\draw (-5.5,.8) node {\includegraphics[width=0.55\textwidth]{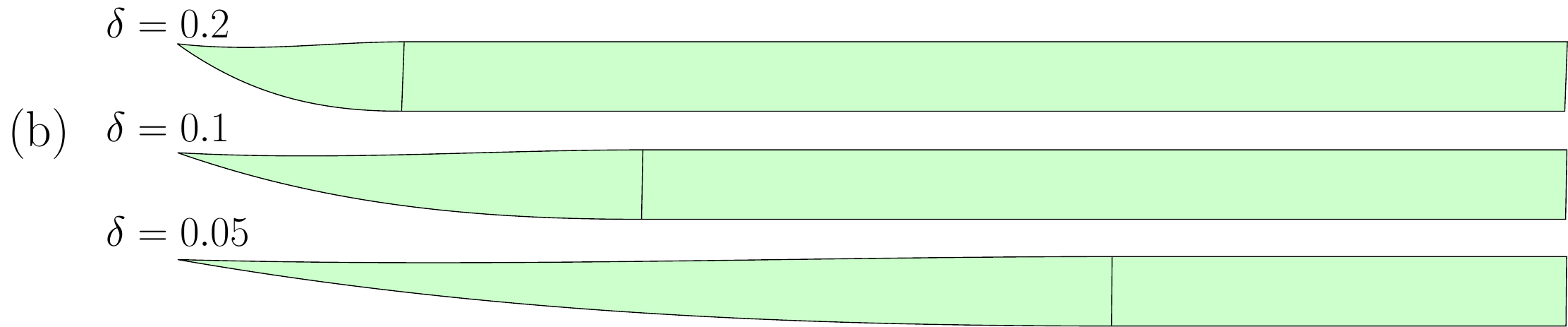}};
\draw (-5.5,3.2) node {\includegraphics[width=0.55\textwidth]{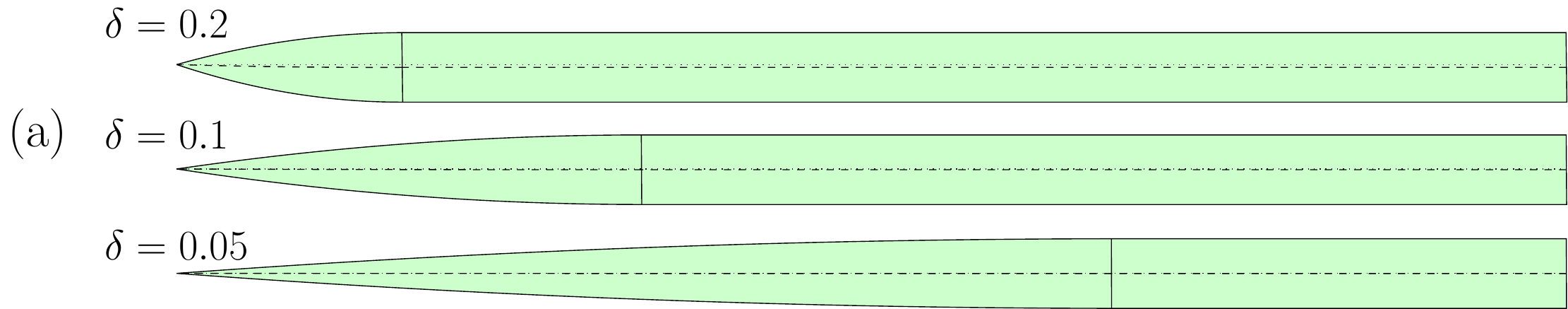}};
\end{tikzpicture}
\end{center}
\caption{Optimal needle shape for different values of $\delta$
in the (a) reference and (b) deformed configuration, scaled for clarity by a factor of 4 vertically as in Figure~\ref{fig:result:stdpars} (b) and (c).
The right panel shows the needle length $L^n$ as a function of $\delta$ (the filled circles mark the configurations on the left).
The black line shows the slope of $1/\delta$.}
\label{fig:nonlineardelta}
\end{figure}
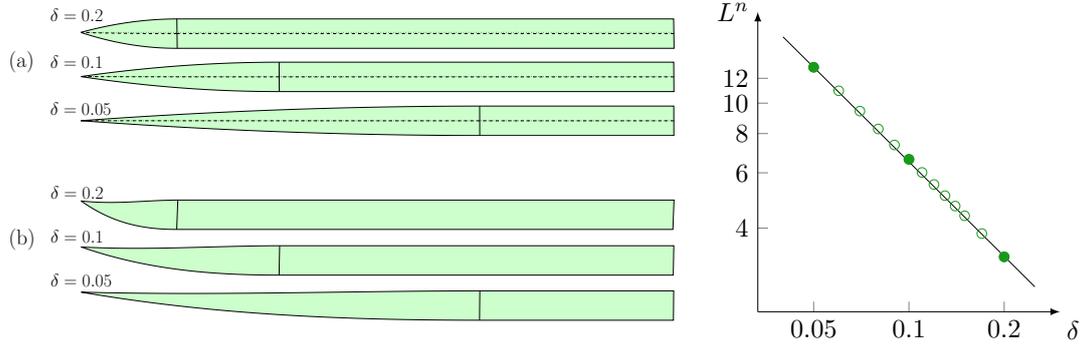

\begin{figure}[ht]
\begin{center}
\begin{tikzpicture}
\begin{scope}[scale=0.8]
\begin{loglogaxis}[height=6cm,width=6cm,log x ticks with fixed point,xtick={3,4,6,10},ytick={2e-3,4e-3,8e-3},yticklabels={$5 \cdot 10^{-4}$,$ 10^{-3}$,  $2 \cdot 10^{-3}$},axis x line*=bottom,axis y line*=left,axis line style={-latex},xlabel={$L^n$},ylabel={$E$},xlabel style={at={(rel axis cs:1,0)}},ylabel style={at={(rel axis cs:0,1)},rotate=-90}]
\addplot[domain=2.4:12.5]{2.73870874e-03/(x/10)};
\addplot[mark=o,color=varafull,only marks] table {
03  8.94808088e-03 
04  6.76004097e-03  
05  5.43130731e-03  
06  4.53894523e-03  
07  3.89836784e-03  
08  3.41620540e-03 
09  3.04017069e-03  
10  2.73870874e-03  
};
\end{loglogaxis}
 \end{scope}
\end{tikzpicture}
\end{center}
\caption{Linearized elastic energy of the optimal needle shape for different values of $L^n$, demonstrating that there is no fixed length scale in the linearized case. The black line shows the slope of $1/L^n$.}
\label{fig:linear}
\end{figure}
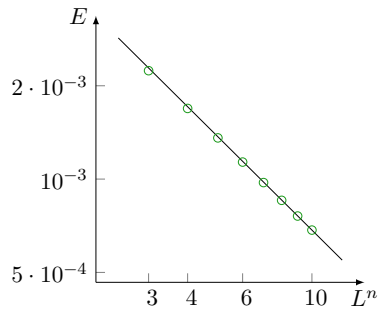

\paragraph*{Linearized elasticity.}
We also considered a linearization of the present model, in which the energy density defined in 
(\ref{eqdefendensity}) is replaced by $W_\mathrm{Lin}(\epsilon):=\frac12 d_{11}(\epsilon_{11}^2+\epsilon_{22}^2) +d_{12}\epsilon_{11}\epsilon_{22}
+2d_{44}\epsilon_{12}^2$ and $W_\mathrm{Lin}^{\vara,\varb}(\epsilon):=W_\mathrm{Lin}(\epsilon\pm\frac12 \delta(e_1\otimes e_2+e_2\otimes e_1))$, with $\epsilon(\deform):=\frac12(D\deform+D\deform^T)-\Id$, and a corresponding linearization of the boundary conditions.
Here $d_{11}:=c_{11}-c_{12}^2/c_{11}$, $d_{12}:=c_{12}-c_{12}^2/c_{11}$  and $d_{44}:=c_{44}$ are the effective elastic coefficients for a two-dimensional plane stress reduction of a  cubic material with elastic constants $c_{11}$, $c_{12}$, $c_{14}$.
The numerics (cf.~Fig.~\ref{fig:linear}) show that the stored elastic energy, after optimizing with respect to the remaining degrees of freedom, decays for increasing $L^n$. In fact, the elastic energy appears to be proportional to $1/{L^n}$, so that the optimal value of $L^n$ seems to be infinity.

\paragraph*{Nonsymmetric, curved needle geometry.}
The optimal needle geometry appears almost, but not exactly, symmetric. In fact, the needle tip is 
shifted upwards by $\approx 0.003$ with respect to
the centerline of the corresponding laminate layer in the reference configuration. Closer inspecting the optimized, parametric curves $\gamma^t$ and $\gamma^b$, one observes a tangential alignment of the needle geometry and the corresponding martensitic laminate layer of constant thickness at the right boundary. For our finest mesh computation the slope of $\gamma^t$ is $1.9\cdot10^{-4}$ and the slope of $\gamma^b$ is $5.4\cdot10^{-5}$ at the point where the needle merges into the laminate (in the reference configuration).

\section{Numerical scheme for the optimization of the needle geometry}
\label{sec:opt}
\begin{figure}[ht]
\begin{center}
\includegraphics[width=0.9\textwidth]{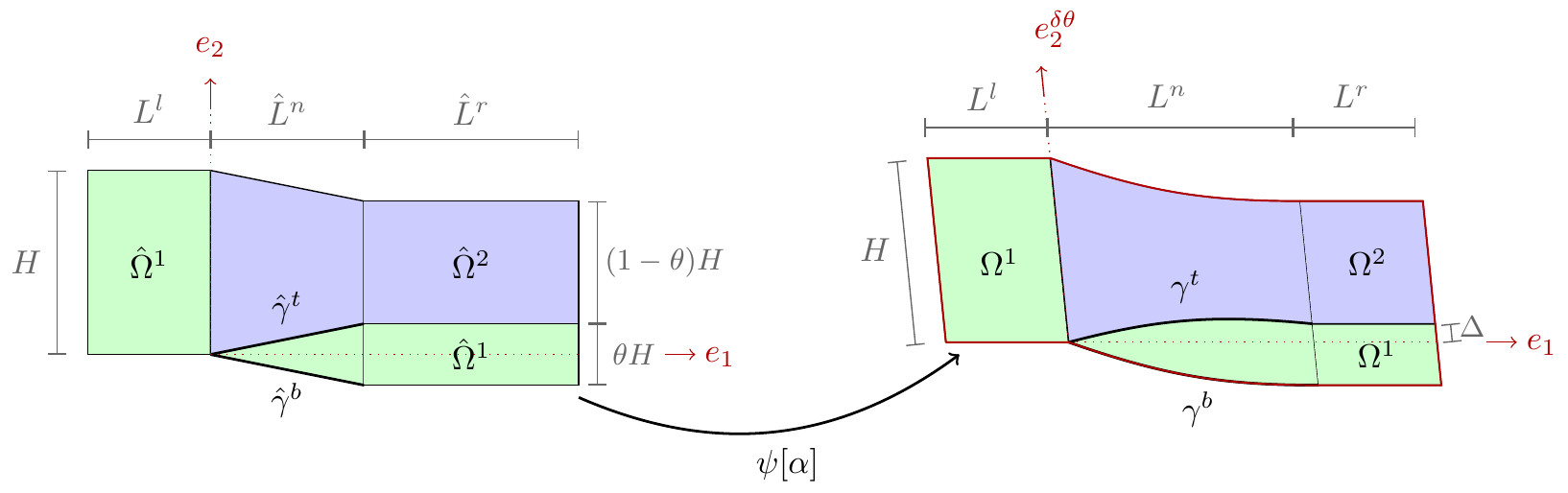}
\end{center}
\caption{Sketch of the transformation of the fundamental cell from the computational domain to the physical reference configuration, cf.~Fig.~\ref{figureneedle1}.}
\label{fig:sketch}
\end{figure}

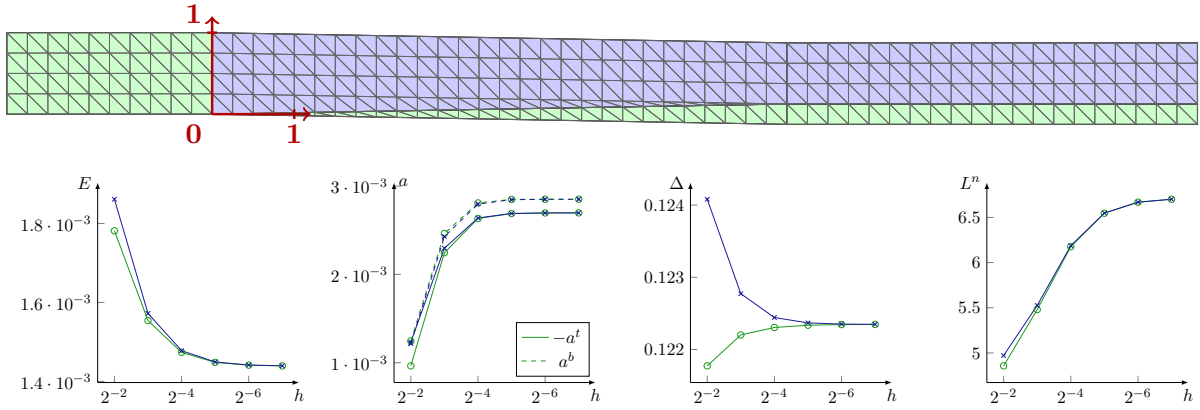
\begin{figure}[ht]
\begin{center}
\includegraphics[width=0.98\textwidth]{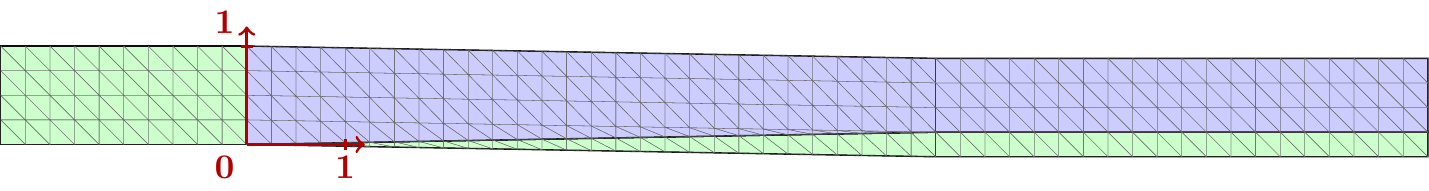}
\medskip

\begin{tikzpicture}[scale=0.6]
\begin{axis}[height=6cm,width=6cm,scaled y ticks=false,xtick={0,2,4},xticklabels={$2^{-2}$,$2^{-4}$,$2^{-6}$},axis x line*=bottom,axis y line*=left,axis line style={-latex},xlabel={$h$},ylabel={$E$},xlabel style={at={(rel axis cs:1,0)}},ylabel style={at={(rel axis cs:0,1)},rotate=-90}]
\addplot[mark=o,color=varafull] table {
0       1.78141863e-03  -9.64618073e-04 1.24588198e-03  1.21773323e-01  4.85702526e+00  -2.16303211e-07 1.05410444e-06  9.79660654e-08  -7.75993738e-07 53        -0.05075304     -2.90544184    
1       1.55405724e-03  -2.24589795e-03 2.46357149e-03  1.22201220e-01  5.48035671e+00  7.02282064e-07  -8.21321953e-07 8.72758583e-07  2.97652637e-07  172       -0.05200104     -2.97675901    
2       1.47349580e-03  -2.63295846e-03 2.80954294e-03  1.22304937e-01  6.17626918e+00  -2.55238699e-07 -3.54097903e-07 -4.73658015e-07 -3.16698173e-07 668       -0.05434760     -3.11082774    
3       1.44822335e-03  -2.68925328e-03 2.84941383e-03  1.22336127e-01  6.54624664e+00  2.40947528e-07  1.15495179e-07  4.44912258e-09  2.76870735e-08  3434      -0.05898801     -3.37585209    
4       1.44121279e-03  -2.69588283e-03 2.85115312e-03  1.22344939e-01  6.66812602e+00  -2.24352307e-07 1.54721000e-07  -4.29457204e-07 2.18929213e-07  42677     -0.06988744     -3.99775531    
5       1.43937199e-03  -2.69705244e-03 2.85108919e-03  1.22347082e-01  6.70095118e+00  1.51835844e-07  -1.34198293e-07 -8.95256409e-08 -2.30055337e-07 450880    -0.10727292     -6.12287112    
};
\addplot[mark=x,color=varbfull] table {
0       1.86135722e-03  -1.23322079e-03 1.21733888e-03  1.24081009e-01  4.97146893e+00  -1.10739582e-06 9.94641177e-07  1.30743909e-06  6.01298867e-07  32        -0.04951914     -2.83492219    
1       1.57224463e-03  -2.29638977e-03 2.42830206e-03  1.22773019e-01  5.52602163e+00  1.17110722e-08  7.20805911e-07  -8.33291582e-07 -7.07475932e-08 164       -0.05159270     -2.95342534    
2       1.47788816e-03  -2.63915666e-03 2.79681566e-03  1.22442909e-01  6.18723908e+00  -2.05153723e-07 7.78889788e-07  1.02860520e-07  1.38055657e-07  514       -0.05419227     -3.10195437    
3       1.44932987e-03  -2.69024744e-03 2.84622198e-03  1.22368567e-01  6.54789611e+00  6.26902414e-08  2.47666642e-08  2.63230056e-08  -5.70083922e-08 4521      -0.05888969     -3.37023828    
4       1.44149438e-03  -2.69606523e-03 2.85043327e-03  1.22352153e-01  6.66829725e+00  -2.32410987e-07 3.32084869e-07  -3.05244055e-07 3.89030479e-07  36185     -0.06978537     -3.99193559    
5       1.43944376e-03  -2.69708729e-03 2.85092028e-03  1.22348719e-01  6.70096043e+00  9.35885787e-08  -6.52263817e-08 7.78719187e-08  2.07102094e-08  431319    -0.10709311     -6.11268585    
};
\end{axis}
\begin{axis}[xshift=6.5cm,height=6cm,width=6cm,scaled y ticks=false,xtick={0,2,4},xticklabels={$2^{-2}$,$2^{-4}$,$2^{-6}$},axis x line*=bottom,axis y line*=left,axis line style={-latex},xlabel={$h$},ylabel={$a$},xlabel style={at={(rel axis cs:1,0)}},ylabel style={at={(rel axis cs:0.1,1)},rotate=-90},legend pos=south east]
\addplot[sharp plot,color=varafull] table[y index=2] {
0       1.78141863e-03  9.64618073e-04 1.24588198e-03  1.21773323e-01  4.85702526e+00  -2.16303211e-07 1.05410444e-06  9.79660654e-08  -7.75993738e-07 53        -0.05075304     -2.90544184    
1       1.55405724e-03  2.24589795e-03 2.46357149e-03  1.22201220e-01  5.48035671e+00  7.02282064e-07  -8.21321953e-07 8.72758583e-07  2.97652637e-07  172       -0.05200104     -2.97675901    
2       1.47349580e-03  2.63295846e-03 2.80954294e-03  1.22304937e-01  6.17626918e+00  -2.55238699e-07 -3.54097903e-07 -4.73658015e-07 -3.16698173e-07 668       -0.05434760     -3.11082774    
3       1.44822335e-03  2.68925328e-03 2.84941383e-03  1.22336127e-01  6.54624664e+00  2.40947528e-07  1.15495179e-07  4.44912258e-09  2.76870735e-08  3434      -0.05898801     -3.37585209    
4       1.44121279e-03  2.69588283e-03 2.85115312e-03  1.22344939e-01  6.66812602e+00  -2.24352307e-07 1.54721000e-07  -4.29457204e-07 2.18929213e-07  42677     -0.06988744     -3.99775531    
5       1.43937199e-03  2.69705244e-03 2.85108919e-03  1.22347082e-01  6.70095118e+00  1.51835844e-07  -1.34198293e-07 -8.95256409e-08 -2.30055337e-07 450880    -0.10727292     -6.12287112    
};
\addlegendentry{$-a^t$}
\addplot[only marks,mark=o,color=varafull,forget plot] table[y index=2] {
0       1.78141863e-03  9.64618073e-04 1.24588198e-03  1.21773323e-01  4.85702526e+00  -2.16303211e-07 1.05410444e-06  9.79660654e-08  -7.75993738e-07 53        -0.05075304     -2.90544184    
1       1.55405724e-03  2.24589795e-03 2.46357149e-03  1.22201220e-01  5.48035671e+00  7.02282064e-07  -8.21321953e-07 8.72758583e-07  2.97652637e-07  172       -0.05200104     -2.97675901    
2       1.47349580e-03  2.63295846e-03 2.80954294e-03  1.22304937e-01  6.17626918e+00  -2.55238699e-07 -3.54097903e-07 -4.73658015e-07 -3.16698173e-07 668       -0.05434760     -3.11082774    
3       1.44822335e-03  2.68925328e-03 2.84941383e-03  1.22336127e-01  6.54624664e+00  2.40947528e-07  1.15495179e-07  4.44912258e-09  2.76870735e-08  3434      -0.05898801     -3.37585209    
4       1.44121279e-03  2.69588283e-03 2.85115312e-03  1.22344939e-01  6.66812602e+00  -2.24352307e-07 1.54721000e-07  -4.29457204e-07 2.18929213e-07  42677     -0.06988744     -3.99775531    
5       1.43937199e-03  2.69705244e-03 2.85108919e-03  1.22347082e-01  6.70095118e+00  1.51835844e-07  -1.34198293e-07 -8.95256409e-08 -2.30055337e-07 450880    -0.10727292     -6.12287112    
};
\addplot[mark=x,color=varbfull,forget plot] table[y index=2] {
0       1.86135722e-03  1.23322079e-03 1.21733888e-03  1.24081009e-01  4.97146893e+00  -1.10739582e-06 9.94641177e-07  1.30743909e-06  6.01298867e-07  32        -0.04951914     -2.83492219    
1       1.57224463e-03  2.29638977e-03 2.42830206e-03  1.22773019e-01  5.52602163e+00  1.17110722e-08  7.20805911e-07  -8.33291582e-07 -7.07475932e-08 164       -0.05159270     -2.95342534    
2       1.47788816e-03  2.63915666e-03 2.79681566e-03  1.22442909e-01  6.18723908e+00  -2.05153723e-07 7.78889788e-07  1.02860520e-07  1.38055657e-07  514       -0.05419227     -3.10195437    
3       1.44932987e-03  2.69024744e-03 2.84622198e-03  1.22368567e-01  6.54789611e+00  6.26902414e-08  2.47666642e-08  2.63230056e-08  -5.70083922e-08 4521      -0.05888969     -3.37023828    
4       1.44149438e-03  2.69606523e-03 2.85043327e-03  1.22352153e-01  6.66829725e+00  -2.32410987e-07 3.32084869e-07  -3.05244055e-07 3.89030479e-07  36185     -0.06978537     -3.99193559    
5       1.43944376e-03  2.69708729e-03 2.85092028e-03  1.22348719e-01  6.70096043e+00  9.35885787e-08  -6.52263817e-08 7.78719187e-08  2.07102094e-08  431319    -0.10709311     -6.11268585    
};
\addplot[dashed,sharp plot,color=varafull] table[y index=3] {
0       1.78141863e-03  9.64618073e-04 1.24588198e-03  1.21773323e-01  4.85702526e+00  -2.16303211e-07 1.05410444e-06  9.79660654e-08  -7.75993738e-07 53        -0.05075304     -2.90544184    
1       1.55405724e-03  2.24589795e-03 2.46357149e-03  1.22201220e-01  5.48035671e+00  7.02282064e-07  -8.21321953e-07 8.72758583e-07  2.97652637e-07  172       -0.05200104     -2.97675901    
2       1.47349580e-03  2.63295846e-03 2.80954294e-03  1.22304937e-01  6.17626918e+00  -2.55238699e-07 -3.54097903e-07 -4.73658015e-07 -3.16698173e-07 668       -0.05434760     -3.11082774    
3       1.44822335e-03  2.68925328e-03 2.84941383e-03  1.22336127e-01  6.54624664e+00  2.40947528e-07  1.15495179e-07  4.44912258e-09  2.76870735e-08  3434      -0.05898801     -3.37585209    
4       1.44121279e-03  2.69588283e-03 2.85115312e-03  1.22344939e-01  6.66812602e+00  -2.24352307e-07 1.54721000e-07  -4.29457204e-07 2.18929213e-07  42677     -0.06988744     -3.99775531    
5       1.43937199e-03  2.69705244e-03 2.85108919e-03  1.22347082e-01  6.70095118e+00  1.51835844e-07  -1.34198293e-07 -8.95256409e-08 -2.30055337e-07 450880    -0.10727292     -6.12287112    
};
\addlegendentry{$a^b$}
\addplot[dashed,mark=x,color=varbfull,forget plot] table[y index=3] {
0       1.86135722e-03  1.23322079e-03 1.21733888e-03  1.24081009e-01  4.97146893e+00  -1.10739582e-06 9.94641177e-07  1.30743909e-06  6.01298867e-07  32        -0.04951914     -2.83492219    
1       1.57224463e-03  2.29638977e-03 2.42830206e-03  1.22773019e-01  5.52602163e+00  1.17110722e-08  7.20805911e-07  -8.33291582e-07 -7.07475932e-08 164       -0.05159270     -2.95342534    
2       1.47788816e-03  2.63915666e-03 2.79681566e-03  1.22442909e-01  6.18723908e+00  -2.05153723e-07 7.78889788e-07  1.02860520e-07  1.38055657e-07  514       -0.05419227     -3.10195437    
3       1.44932987e-03  2.69024744e-03 2.84622198e-03  1.22368567e-01  6.54789611e+00  6.26902414e-08  2.47666642e-08  2.63230056e-08  -5.70083922e-08 4521      -0.05888969     -3.37023828    
4       1.44149438e-03  2.69606523e-03 2.85043327e-03  1.22352153e-01  6.66829725e+00  -2.32410987e-07 3.32084869e-07  -3.05244055e-07 3.89030479e-07  36185     -0.06978537     -3.99193559    
5       1.43944376e-03  2.69708729e-03 2.85092028e-03  1.22348719e-01  6.70096043e+00  9.35885787e-08  -6.52263817e-08 7.78719187e-08  2.07102094e-08  431319    -0.10709311     -6.11268585    
};
\addplot[only marks,mark=o,color=varafull,forget plot] table[y index=3] {
0       1.78141863e-03  9.64618073e-04 1.24588198e-03  1.21773323e-01  4.85702526e+00  -2.16303211e-07 1.05410444e-06  9.79660654e-08  -7.75993738e-07 53        -0.05075304     -2.90544184    
1       1.55405724e-03  2.24589795e-03 2.46357149e-03  1.22201220e-01  5.48035671e+00  7.02282064e-07  -8.21321953e-07 8.72758583e-07  2.97652637e-07  172       -0.05200104     -2.97675901    
2       1.47349580e-03  2.63295846e-03 2.80954294e-03  1.22304937e-01  6.17626918e+00  -2.55238699e-07 -3.54097903e-07 -4.73658015e-07 -3.16698173e-07 668       -0.05434760     -3.11082774    
3       1.44822335e-03  2.68925328e-03 2.84941383e-03  1.22336127e-01  6.54624664e+00  2.40947528e-07  1.15495179e-07  4.44912258e-09  2.76870735e-08  3434      -0.05898801     -3.37585209    
4       1.44121279e-03  2.69588283e-03 2.85115312e-03  1.22344939e-01  6.66812602e+00  -2.24352307e-07 1.54721000e-07  -4.29457204e-07 2.18929213e-07  42677     -0.06988744     -3.99775531    
5       1.43937199e-03  2.69705244e-03 2.85108919e-03  1.22347082e-01  6.70095118e+00  1.51835844e-07  -1.34198293e-07 -8.95256409e-08 -2.30055337e-07 450880    -0.10727292     -6.12287112    
};
\addplot[only marks,mark=x,color=varbfull,forget plot] table[y index=3] {
0       1.86135722e-03  1.23322079e-03 1.21733888e-03  1.24081009e-01  4.97146893e+00  -1.10739582e-06 9.94641177e-07  1.30743909e-06  6.01298867e-07  32        -0.04951914     -2.83492219    
1       1.57224463e-03  2.29638977e-03 2.42830206e-03  1.22773019e-01  5.52602163e+00  1.17110722e-08  7.20805911e-07  -8.33291582e-07 -7.07475932e-08 164       -0.05159270     -2.95342534    
2       1.47788816e-03  2.63915666e-03 2.79681566e-03  1.22442909e-01  6.18723908e+00  -2.05153723e-07 7.78889788e-07  1.02860520e-07  1.38055657e-07  514       -0.05419227     -3.10195437    
3       1.44932987e-03  2.69024744e-03 2.84622198e-03  1.22368567e-01  6.54789611e+00  6.26902414e-08  2.47666642e-08  2.63230056e-08  -5.70083922e-08 4521      -0.05888969     -3.37023828    
4       1.44149438e-03  2.69606523e-03 2.85043327e-03  1.22352153e-01  6.66829725e+00  -2.32410987e-07 3.32084869e-07  -3.05244055e-07 3.89030479e-07  36185     -0.06978537     -3.99193559    
5       1.43944376e-03  2.69708729e-03 2.85092028e-03  1.22348719e-01  6.70096043e+00  9.35885787e-08  -6.52263817e-08 7.78719187e-08  2.07102094e-08  431319    -0.10709311     -6.11268585    
};
\end{axis}
\begin{axis}[xshift=13cm,height=6cm,width=6cm,scaled y ticks=false,xtick={0,2,4},xticklabels={$2^{-2}$,$2^{-4}$,$2^{-6}$},axis x line*=bottom,axis y line*=left,axis line style={-latex},xlabel={$h$},ylabel={$\Delta$},xlabel style={at={(rel axis cs:1,0)}},ylabel style={at={(rel axis cs:0,1)},rotate=-90},y tick label style={/pgf/number format/.cd,fixed,fixed zerofill,precision=3}]
\addplot[mark=o,color=varafull] table[y index=4] {
0       1.78141863e-03  -9.64618073e-04 1.24588198e-03  1.21773323e-01  4.85702526e+00  -2.16303211e-07 1.05410444e-06  9.79660654e-08  -7.75993738e-07 53        -0.05075304     -2.90544184    
1       1.55405724e-03  -2.24589795e-03 2.46357149e-03  1.22201220e-01  5.48035671e+00  7.02282064e-07  -8.21321953e-07 8.72758583e-07  2.97652637e-07  172       -0.05200104     -2.97675901    
2       1.47349580e-03  -2.63295846e-03 2.80954294e-03  1.22304937e-01  6.17626918e+00  -2.55238699e-07 -3.54097903e-07 -4.73658015e-07 -3.16698173e-07 668       -0.05434760     -3.11082774    
3       1.44822335e-03  -2.68925328e-03 2.84941383e-03  1.22336127e-01  6.54624664e+00  2.40947528e-07  1.15495179e-07  4.44912258e-09  2.76870735e-08  3434      -0.05898801     -3.37585209    
4       1.44121279e-03  -2.69588283e-03 2.85115312e-03  1.22344939e-01  6.66812602e+00  -2.24352307e-07 1.54721000e-07  -4.29457204e-07 2.18929213e-07  42677     -0.06988744     -3.99775531    
5       1.43937199e-03  -2.69705244e-03 2.85108919e-03  1.22347082e-01  6.70095118e+00  1.51835844e-07  -1.34198293e-07 -8.95256409e-08 -2.30055337e-07 450880    -0.10727292     -6.12287112    
};
\addplot[mark=x,color=varbfull] table[y index=4] {
0       1.86135722e-03  -1.23322079e-03 1.21733888e-03  1.24081009e-01  4.97146893e+00  -1.10739582e-06 9.94641177e-07  1.30743909e-06  6.01298867e-07  32        -0.04951914     -2.83492219    
1       1.57224463e-03  -2.29638977e-03 2.42830206e-03  1.22773019e-01  5.52602163e+00  1.17110722e-08  7.20805911e-07  -8.33291582e-07 -7.07475932e-08 164       -0.05159270     -2.95342534    
2       1.47788816e-03  -2.63915666e-03 2.79681566e-03  1.22442909e-01  6.18723908e+00  -2.05153723e-07 7.78889788e-07  1.02860520e-07  1.38055657e-07  514       -0.05419227     -3.10195437    
3       1.44932987e-03  -2.69024744e-03 2.84622198e-03  1.22368567e-01  6.54789611e+00  6.26902414e-08  2.47666642e-08  2.63230056e-08  -5.70083922e-08 4521      -0.05888969     -3.37023828    
4       1.44149438e-03  -2.69606523e-03 2.85043327e-03  1.22352153e-01  6.66829725e+00  -2.32410987e-07 3.32084869e-07  -3.05244055e-07 3.89030479e-07  36185     -0.06978537     -3.99193559    
5       1.43944376e-03  -2.69708729e-03 2.85092028e-03  1.22348719e-01  6.70096043e+00  9.35885787e-08  -6.52263817e-08 7.78719187e-08  2.07102094e-08  431319    -0.10709311     -6.11268585    
};
\end{axis}
\begin{axis}[xshift=19.5cm,height=6cm,width=6cm,scaled y ticks=false,xtick={0,2,4},xticklabels={$2^{-2}$,$2^{-4}$,$2^{-6}$},axis x line*=bottom,axis y line*=left,axis line style={-latex},xlabel={$h$},ylabel={$L^n$},xlabel style={at={(rel axis cs:1,0)}},ylabel style={at={(rel axis cs:0,1)},rotate=-90}]
\addplot[mark=o,color=varafull] table[y index=5] {
0       1.78141863e-03  -9.64618073e-04 1.24588198e-03  1.21773323e-01  4.85702526e+00  -2.16303211e-07 1.05410444e-06  9.79660654e-08  -7.75993738e-07 53        -0.05075304     -2.90544184    
1       1.55405724e-03  -2.24589795e-03 2.46357149e-03  1.22201220e-01  5.48035671e+00  7.02282064e-07  -8.21321953e-07 8.72758583e-07  2.97652637e-07  172       -0.05200104     -2.97675901    
2       1.47349580e-03  -2.63295846e-03 2.80954294e-03  1.22304937e-01  6.17626918e+00  -2.55238699e-07 -3.54097903e-07 -4.73658015e-07 -3.16698173e-07 668       -0.05434760     -3.11082774    
3       1.44822335e-03  -2.68925328e-03 2.84941383e-03  1.22336127e-01  6.54624664e+00  2.40947528e-07  1.15495179e-07  4.44912258e-09  2.76870735e-08  3434      -0.05898801     -3.37585209    
4       1.44121279e-03  -2.69588283e-03 2.85115312e-03  1.22344939e-01  6.66812602e+00  -2.24352307e-07 1.54721000e-07  -4.29457204e-07 2.18929213e-07  42677     -0.06988744     -3.99775531    
5       1.43937199e-03  -2.69705244e-03 2.85108919e-03  1.22347082e-01  6.70095118e+00  1.51835844e-07  -1.34198293e-07 -8.95256409e-08 -2.30055337e-07 450880    -0.10727292     -6.12287112    
};
\addplot[mark=x,color=varbfull] table[y index=5] {
0       1.86135722e-03  -1.23322079e-03 1.21733888e-03  1.24081009e-01  4.97146893e+00  -1.10739582e-06 9.94641177e-07  1.30743909e-06  6.01298867e-07  32        -0.04951914     -2.83492219    
1       1.57224463e-03  -2.29638977e-03 2.42830206e-03  1.22773019e-01  5.52602163e+00  1.17110722e-08  7.20805911e-07  -8.33291582e-07 -7.07475932e-08 164       -0.05159270     -2.95342534    
2       1.47788816e-03  -2.63915666e-03 2.79681566e-03  1.22442909e-01  6.18723908e+00  -2.05153723e-07 7.78889788e-07  1.02860520e-07  1.38055657e-07  514       -0.05419227     -3.10195437    
3       1.44932987e-03  -2.69024744e-03 2.84622198e-03  1.22368567e-01  6.54789611e+00  6.26902414e-08  2.47666642e-08  2.63230056e-08  -5.70083922e-08 4521      -0.05888969     -3.37023828    
4       1.44149438e-03  -2.69606523e-03 2.85043327e-03  1.22352153e-01  6.66829725e+00  -2.32410987e-07 3.32084869e-07  -3.05244055e-07 3.89030479e-07  36185     -0.06978537     -3.99193559    
5       1.43944376e-03  -2.69708729e-03 2.85092028e-03  1.22348719e-01  6.70096043e+00  9.35885787e-08  -6.52263817e-08 7.78719187e-08  2.07102094e-08  431319    -0.10709311     -6.11268585    
};
\end{axis}

\end{tikzpicture}

\end{center}
\caption{Top: Coarsest finite element mesh, $h=2^{-2}$, \criss{} type. Bottom:  Convergence plots for the total energy and the four optimization parameters (the two curvatures $-a^t$ and $a^b$ in the second plot). In each plot the results for \criss{} (blue x markers, corresponding to the mesh depicted on top) and \cross{} (green circle markers) grids are shown for successive refinement by quadrisection. A symmetric needle would have $\Delta=0.125$.}
\label{fig:convergence}
\end{figure}

In order to numerically optimize over the geometry we use a change-of-variables technique. Precisely, we let
the physical domain $\trfdom$ be the image of a (fixed) computational domain $\refdom$  via a map
$\transform[\alpha]$ depending on a finite set of design parameters $\param$.
The physical reference configuration represents the undeformed state of the fundamental cell 
of the martensitic twin microstructure and is deformed by the actual elastic deformation $\phi$.
Computational domain and physical reference configuration are sketched in Figure~\ref{fig:sketch}. 
In the physical reference configuration the top and bottom boundaries of the needle are described by curves $\gamma^t$ and $\gamma^b$, which in turn are parametrized via $\transform[\alpha]$ 
over two (fixed) piecewise affine curves $\hat \gamma^t$ and $\hat \gamma^b$ at corresponding positions in the computational domain.
The horizontal length $L$ of the fundamental cell is fixed and splits up into the length of left $\eigenstrain^\vara$ variant $L^l$, which is fixed as well, the length of the needle tip $L^n$ and the length of the laminate component on the right $L^r$ (ideally this is an infinite domain, we make it finite only for computational reasons).
The needle length $L^n$ in the physical reference configuration is one of the design parameters and consequently $L^r = L-L^l-L^n$.
The overall height for the fundamental cell is denoted by $H$, which splits up into fractions $\theta$ and $(1-\theta)$  corresponding to the two martensitic variants. In our computations we chose $H=1$. The values of $L$ and $L^l$ were chosen depending on the expected range of $L^n$, i.e. $L=14.5$, $L^l=2.5$ for Fig.~\ref{fig:result:stdpars} and Fig.~\ref{fig:convergence}, $L=22.5$, $L^l=2.5$ for Fig.~\ref{fig:nonlineardelta}, and $L=16$, $L^l=2$ for Fig.~\ref{fig:linear}. 

Another design parameter  is the width $\Delta$ of the needle cross section at the end of the needle tip above the $x_1$ axis along the sheared direction $e^{\delta \theta}_2$.
Correspondingly, the width below the $x_1$ axis is $\theta H-\Delta$.
The parametrization $\transform[\alpha]$ is composed of three transformations:
(1) a linear stretching of the needle interval along the $x_1$ axis by a factor $\frac{L^n}{\hat L^n}$, 
(2) a piecewise affine deformation on slices in the direction of $e_2$ within the needle interval $[0,L^n]$ which 
maps $\hat \gamma^b$ to  $\gamma^b$, $\hat \gamma^t$ to  $\gamma^t$, and $\hat \gamma^b + H e_2$ to  $\gamma^b+ H e_2$, and 
(3) a shearing deformation keeping $e_1$ fixed and mapping $e_2$ onto $e_2^{\delta \theta}$.
To describe the curved needle geometries we consider simple quadratic curves $\gamma^{t} = \{ \left(x,a^{t} x^2 + (\frac{\Delta}{L^n} - a^{t} L^n) x \right)\,:\, x \in [0,L^n]\}$, 
$\gamma^{b} = \{ \left(x,a^{b} x^2 + (\frac{\Delta-\theta H}{L^n} - a^{b} L^n) x \right)\,:\, x \in [0,L^n]\}$
where the coefficients $a^{t}$ and $a^{b}$ are additional degrees of freedom. 
Altogether we consider as the vector of design variables $\alpha := (\Delta, L^n, a^b, a^t)$.

Given this geometric configuration we aim at minimizing the shape functional $\param \mapsto \cost[\param]$, where 
$\cost[\param] := \energy[\alpha, \deform[\param]]$ with $\deform[\param]$ being the elastic deformation minimizing the stored elastic energy (\ref{eqdefetot}),
which we assume to be unique.
Taking into account the parametrizations $\trfdom^\varboth := \transform[\param](\refdom^\varboth)$  
of the two physical  domains  $\Omega^\vara$ and $\Omega^\varb$, we obtain for the stored elastic energy 
for a deformation $\hat \phi := \phi \circ \transform[\alpha]$
\begin{eqnarray*} 
{\refenergy} [\param, \refdeform] &:=& E[\deform,\Omega^\vara, \Omega^\varb]  =
\int_{\transform[\alpha](\refdom^\vara)} \edensity^\vara(\D \deform) \d x + 
\int_{\transform[\alpha](\refdom^\varb)} \edensity^\varb(\D \deform) \d x \\
&=& \sum_{m=\vara,\,\varb} \int_{\refdom^m} 
\edensity^m\left(\Dref \refdeform(\xref) \left(\Dref \psi[\param](\xref)\right)^{-1}\right)
\det \Dref \psi[\param](\xref) \d \xref\,.
\end{eqnarray*} 
Here, we assume that $\transform[\alpha]$ is bijective and use 
the chain rule $\D \deform(x) = \Dref \refdeform[\param](\xref) \left(\Dref \psi[\param](\xref)\right)^{-1}$
where $\Dref$ is the differential in reference coordinates $\xref$ with $x=\transform[\alpha](\xref)$.
Finally, we consider periodic boundary conditions on the top and bottom boundary of our fundamental domain
and natural boundary conditions on the left and on the right.

A necessary condition for a reference deformation $\refdeform$ to minimize the elastic energy 
$\refenergy[\param, \refdeform]$ for fixed $\param$ is given by the Euler-Lagrange equation 
$0 = \refenergy_{,\refdeform} [\param, \refdeform]\,,$
where $\refenergy_{,\refdeform}$ denotes the Fr{\'e}chet derivative with respect to $\refdeform$.
For the spatial discretization
we consider an admissible simplicial finite element mesh for the reference domains $\refdom^{1,2}$
with coinciding nodes on the common interface. Denoting by $\hat V_h$ the associated space of piecewise affine and continuous functions from $\refdom$ to $\R^2$
we consider the conforming finite element discretization of our variational problem with $\refdeform$ in $\hat V_h$.
The solution $\refdeform[\param]$ is computed using Newton's scheme for the state equation $\energy_{,\refdeform}[\param, \refdeform] = 0$ where a small number of iterations (up to 11) are sufficient to obtain a residual in the $L^2$ norm of $10^{-6}$.
This scheme can easily be extended via a suitable projection to incorporate translational invariance 
$\int_{\trfdom   } ( \phi - \Id) \d x =0$.

To minimize the cost functional $\cost[\param] = \refenergy[\param,\refdeform[\param]]$ 
we consider a cascadic nonlinear conjugate gradient scheme.  
For the linesearch with step size control in the descent we proceed as follows.
Given a descent direction $q$ in the design variable $\param$ we define 
$f(t) := \cost(\param+t q)$ and let $p$ be a quadratic polynomial solving the interpolation problem
$p(0) = f(0)$, $p'(0) = f'(0) = \cost_{,\param}[\param](q)$, and $p(\lambda) = f(\lambda)$
for a suitable $\lambda >0$. The minimum of $p$ is attained for the time step 
$\lambda^\ast = - \frac{f'(0) \lambda^2}{2 (f(\lambda)- f'(0) \lambda - f(0))}$. 
 This time step is then used as input for  Armijo's time step control applied to $t \mapsto f(t)$.
We descend until $|\cost_{,\param}[\param]|<10^{-5}$, with at most 101 descent steps on each level of the cascadic scheme in our numerical experiments.


We have experimentally studied the convergence of the elastic energy and the geometry degrees of freedom 
$\alpha$. For the shape  shown in Fig.~\ref{fig:result:stdpars} convergence plots for a spatial grid size $h=2^{-n}$ for $n=2,\ldots 7$ 
are presented in Fig.~\ref{fig:convergence}. As a robustness check we have also considered different types of finite element meshes, i.e. meshes with diagonal faces in the $(1,-1)$ (\criss) and $(1,1)$ (\cross) directions, respectively.
All computations in Sec.~\ref{sec:results} have been performed on a mesh with $h=2^{-6}$ with a \criss{} grid, oriented as in Fig.~\ref{fig:convergence}.

\section*{Acknowledgements}
This work was partially funded by the Deutsche Forschungsgemeinschaft (DFG, German Research Foundation) via Project  211504053 -- SFB 1060.

\bibliography{manuscript}
\bibliographystyle{alpha-noname}

\end{document}